\documentclass{PoS}

\usepackage{epsfig}
\usepackage{amsmath}

\def\mres{m_{\rm res}}
\def\MSbar{{\overline{\mbox{MS}}}}
\def\MSb{{\overline{\mbox{MS}}}}
\def\msbar{{\overline{\mbox{\scriptsize MS}}}}

\def\SMOM{{\rm SMOM}}
\def\SMOMgm{{\rm SMOM}_{\gamma_\mu}}
\def\RIMOM{{\rm RI/MOM}}
\def\RI'MOM{{\rm RI'/MOM}}
\def\RISMOM{{\rm RI/SMOM}}
\def\RISMOMgm{{\rm RI/SMOM}_{\gamma_\mu}}

\def\rimom{\mbox{\scriptsize RI/MOM}}
\def\ri'mom{\mbox{\scriptsize RI'/MOM}}
\def\rismom{\mbox{\scriptsize RI/SMOM}}
\def\rismomgm{\mbox{\scriptsize RI/SMOM}_{\gamma_\mu}}
\def\Tr{\mbox{Tr}}
\def\Proj{{\mathcal P}}

\def\cancel#1#2{\ooalign{$\hfil#1\mkern1mu/\hfil$\crcr$#1#2$}}
\def\slash#1{\mathpalette\cancel{#1}}

\newcounter{Outline}
\title{Non-perturbative renormalization in lattice QCD}

\ShortTitle{Non-perturbative renormalization in lattice QCD}

\author{\speaker{Yasumichi Aoki}\\
        RIKEN BNL Research Center, Brookhaven National Laboratory,
	Upton, NY 11973, USA\\
        E-mail: \email{yaoki@bnl.gov}}

\abstract{
Recent developments in non-perturbative renormalization for lattice 
QCD are reviewed with a particular emphasis on RI/MOM scheme
and its variants, RI/SMOM schemes. Summary of recent developments
in Schr\"odinger functional scheme, as well as the summary of related topics
are presented. Comparison of strong coupling constant and the strange quark
mass from various methods are made.}

\FullConference{The XXVII International Symposium on Lattice Field Theory - LAT2009\\
		 July 26-31 2009\\
		 Peking University, Beijing, China}

\begin{document}

\section{Introduction}

In many applications of lattice QCD it is necessary to deal with
renormalization.  A simplest example is to obtain the quark masses
which are the fundamental parameters in the standard model.  A bare quark mass
is logarithmically divergent in the continuum limit, hence, is not
defined without proper regularization and renormalization.
A less simple example is to calculate a
transition or decay amplitude of hadrons induced by interactions in the
electroweak theory. The low energy effective operators are constructed
through operator product expansion. They, in turn, are plugged in the
QCD Lagrangian, with the Wilson coefficients calculated through
(continuum) perturbation theory. The coefficients are often divergent, thus
should be defined through renormalization with a certain renormalization
scheme and scale.  The lattice part is to calculate the matrix element
of the operator between hadron states. The matrix element of the bare
operator has the same divergence as the Wilson coefficient. In divergent cases,
the matrix element or the operator must undergo the
renormalization on the lattice.  The same renormalization scheme and
scale as the Wilson coefficient must be used to compensate 
the dependence on them,
so that the final, physical matrix element is scheme and scale independent. 

As these typical examples show, there are two necessary steps required 
for the lattice renormalization:
1) removing the ultraviolet divergence
in the observables and 2) matching to the scheme convenient to the
continuum renormalization, such as $\MSbar$.
The step 2) can be performed in continuum perturbation theory (cPT) once
1) is defined in a regularization independent (RI) way.
The lattice perturbation theory (LPT) can be used to perform 1). However,
the convergence of the series in general is not good due to tadpole
contributions. Although a cure is provided by mean field improved
perturbation theory, one has to deal with the ambiguity due to
the choice of mean field coupling. In addition, for some 
quantities the matching 2) has been calculated to three loops in cPT.
Such a calculation cannot be fully exploited if the LPT is used for the
step 1) due to the difficulty of calculating two or more loops in LPT.
These problems can be bypassed by the use of non-perturbative
renormalization (NPR) applied for the step 1).

The NPR on the lattice was born early in the previous decade. There have
been dedicated plenary presentations in this series of conferences
\cite{Rossi:1996yu,Testa:1997ne,Sint:2000vc,Sommer:2002en}.

The popular NPR schemes to date are the momentum subtraction
(RI/MOM) and Schr\"odinger functional (SF) schemes.
The SF scheme often refers to its combined use with step scaling.
The RI/MOM scheme is widely used in the light quark sector
of state-of-the-art lattice calculations today.
The use of SF scheme for light quarks is slightly restricted,
presumably due to the need of tailored gauge configurations and
subtleties when applying to the chiral fermions. 
The SF scheme has been extended to incorporate the heavy quark
effective theory (HQET) on the lattice. It has been discussed
in \cite{Sommer:2002en} and more recently to some detail by 
Della Morte \cite{DellaMorte:2007ny}.

As is well known there is the so called ``window problem'' in the
RI/MOM scheme.  To have minimal effect from the problem
some practical prescription is required for the RI/MOM
scheme.  As an extension of the RI/MOM scheme, the RI/SMOM scheme
was proposed recently and applied to some realistic calculations.
The new scheme is meant to ease the ``window problem'' by squeezing out
unwanted non-perturbative effects from the setup of the scheme.
This review focuses mainly on the recent development in RI/MOM and
RI/SMOM schemes on how to work around the problem.

In what follows, RI/MOM and RI/SMOM schemes are discussed in Sec.~2.
Recent developments in SF scheme are given in Sec.~3. 
Sec.~4 contains application specific developments, such as the
calculation of the strong coupling constant and quark masses.
Concluding remarks are given in Sec.~5.

\section{Developments in the RI/MOM scheme}

The RI/MOM scheme \cite{Martinelli:1993dq,Martinelli:1994ty}
is a popular non-perturbative renormalization scheme
for multi-quark operators today.
It has been applied to virtually all types of light quark 
discretizations used to date. The renormalization condition is
imposed on the forward vertex function of the operator with
external off-shell quark states having momentum $p$. 
This scheme is handy as no special gauge
configurations are required.  A drawback is that there is the so
called window problem, i.e.~the renormalization scale $\mu=\sqrt{p^2}$
needs to be in the range $\Lambda_{QCD}\ll \mu \ll 1/a$.
Finer lattice simulations are less problematic. However for the typical
lattice cutoff used today, this could be a source of a sizable
systematic error. 
In principle, the RI/MOM scheme can be used to determine the coefficient
of the higher dimensional operators that appear in the Symanzik
on-shell improvement. However, more operators have to be introduced
than the SF scheme to properly disentangle the linear relation due to
the use of off-shell external states, which would make the tuning difficult
in practice.  Operators with chiral or twisted mass fermions have
no problem with this point since there are no $O(a)$ operators to mix
at on-shell for typical applications.

Several remarkable developments in the RI/MOM scheme and its extension
are reported by this year. To discuss them it is required to 
have a review of the RI/MOM scheme.

\subsection{RI/MOM formulation}
For illustrative purpose let us go through the RI/MOM renormalization
of a flavor non-singlet quark bilinear operator $O_\Gamma=\overline{u}\Gamma d$.
The concept is easily extended to three, four $\cdots$ quark
operators.
Later on, the pseudoscalar/scalar operator is examined in detail for
the quark mass renormalization.
The bare operator $O_\Gamma$ with the Dirac structure $\Gamma$ is
renormalized with $Z_\Gamma$ as 
\begin{equation}
 O_\Gamma^R = Z_\Gamma O_\Gamma.
\end{equation}
The RI/MOM renormalization condition is imposed to the amputated vertex
function $\Pi_\Gamma(p_1,p_2)$ with Landau-gauge fixed external
mass-less off-shell quark states with the same incoming and out-going
momenta 
$p_1=p_2=p$,
\begin{equation}
 \frac{Z_\Gamma}{Z_q}\Tr[\Pi_\Gamma\cdot\Proj_\Gamma] = 1,\label{eq:ren_cond_Gamma}
\end{equation}
where $Z_q$ is the quark wavefunction renormalization factor, i.e.~
$S^R=Z_qS$ with $S^R$ and $S$ being the renormalized and bare
quark propagator.
The projection operator $\Proj_\Gamma$ is chosen so that the condition
properly removes all the ultraviolet divergence and 
holds trivially at tree level with $Z_\Gamma=Z_q=1$.
The renormalization scale is set through the external off-shell momentum
$\mu^2=p^2$.

There are two different schemes due to slightly different definitions
of the wavefunction renormalization. The RI/MOM $Z_q$ is defined
through the vector vertex with $\Proj_V\propto\gamma_\mu$
\begin{equation}
 \frac{Z_V}{Z_q}\frac{1}{48}\Tr[\Pi_{V_\mu}\cdot\gamma_\mu] =
  1. \label{eq:ren_cond_V}
\end{equation}
In the continuum or by the use of conserved vector current on the lattice,
since $Z_V=1$ this determines $Z_q$.  With the local vector 
current on the lattice $Z_V$ needs to be determine by other means.
In the continuum, this condition is equivalent through the Ward-Takahashi
(WT) identity with the condition on the inverse quark propagator,
\begin{equation}
 \left. \frac{1}{Z_q}\frac{1}{12}\mbox{Tr}\left[-i\frac{\partial}
		       {\partial \slash{p}}S^{-1}(p)\right] \right|_{p^2=\mu^2}
  = 1.\label{eq:RIq}\\
\end{equation}
Another scheme used to date is RI'/MOM scheme where instead
of Eq.~(\ref{eq:ren_cond_V}) or (\ref{eq:RIq}) $Z_q$ is
defined through
\begin{equation}
  \left. \frac{1}{Z_q}\frac{1}{12}\mbox{Tr}
   \left[-i \frac{\slash{p}}{p^2} S^{-1}(p)\right]
  \right|_{p^2=\mu^2} = 1.\label{eq:RI'q}
\end{equation}
These two definitions of $Z_q$ are quite similar, which can
be seen by the fact that the difference only appears at two loops
in continuum perturbation theory \cite{Chetyrkin:1999pq}.
To date RI'/MOM scheme is being used mostly for Wilson type fermions,
while the RI/MOM is being utilized in domain-wall (DW) and overlap fermions. 

Once the $Z_\Gamma$ in RI/MOM or RI'/MOM scheme is obtained,
it can be converted to that in other schemes by multiplying a conversion
factor calculated in perturbation theory.

These are the essence of RI/MOM schemes. 
In the lattice computation, there emerges a practical
issues associated with the use of momentum for the
renormalization scale.
The so-called window problem arises as the renormalization scale
needs to satisfy two conditions. The first condition is
\begin{itemize}
 \item $\mu\gg\Lambda_{\rm QCD}$.
\end{itemize} 
The matching factor / Wilson coefficient is calculated in perturbation
theory. At a given order, the truncation error grows as $\mu$
decreases. Furthermore, if the operator couples to the 
Nambu-Goldtone (NG) boson due to spontaneous symmetry breaking (SSB), the
effect cannot be corrected by perturbation theory and fails the
matching.
The condition is required to have these effects small.
It should be noted, however, for some operators the SSB effect
diverges as $1/m$ which has to be subtracted.
Secondly,
\begin{itemize}
 \item $\mu\ll 1/a$.
\end{itemize}
This is to reduce the discretization error, such like $(pa)^2$.
However, the $(pa)^2\ge 1$  region is used in typical calculations.
It appears, in practice, $\mu\ll \pi/a$ is OK.

These conditions provide the guideline to have the systematic error
small for the RI/MOM schemes and the matching. A more detailed discussion
follows by using an example of quark mass 
renormalization for the estimate of the systematic errors 
with $n_f=2+1$ DWFs. Here for simplicity
we assume exact chiral symmetry. An issue of the small explicit 
chiral symmetry breaking will be discussed in Sec.~\ref{sec:rimom_remarks}.
The mass renormalization for $m^R=Z_m m$ is conveniently calculated
through the relation 
\begin{equation}
 Z_m = 1/Z_S = 1/Z_P\footnote{$Z_m=1/Z_P\ne 1/Z_S$ for Wilson type fermions.},
  \label{eq:ZmZsZp} 
\end{equation}
where $Z_m$, $Z_S$, $Z_P$ are quark mass, flavor non-singlet scalar
and pseudoscalar renormalization factors.
The ratio $Z_q/Z_\Gamma$ is calculated from 
the traced projected vertex functions through the
renormalization condition Eq.~(\ref{eq:ren_cond_Gamma}),
\begin{equation}
 \Lambda_\Gamma = \Tr[\Pi_\Gamma\cdot\Proj_\Gamma] \to \frac{Z_q}{Z_\Gamma}.
\end{equation}
It immediately turns out that $\Lambda_S\ne\Lambda_P$ due to SSB, 
i.e., $\Lambda_P$ diverges as $1/m$ \cite{Aoki:2007xm,Aoki:2009ka}.
Therefore one needs to subtract the divergence or abandon the use of
$\Lambda_P$. 
Here we proceed to take the latter choice and use
\footnote{In quenched approximation
there also be a divergence in the scalar vertex as $1/m^2$
\cite{Blum:2001sr}, which is suppressed 
in dynamical calculation from the quark determinant effect.}
\begin{equation}
 \Lambda_S =  \frac{1}{12}\Tr[\Pi_S\cdot I] \to \frac{Z_q}{Z_S} = Z_q Z_m.
\end{equation}
For Wilson type fermions, the pseudoscalar vertex must be used
for the mass renormalization. The divergent pion pole contribution
must be subtracted \cite{Cudell:1998ic}.  Fitting the quark mass
dependence with NLO OPE expression is one way. 
Or one can construct a object 
free from the pole combining the multi mass points \cite{Giusti:2000jr}.

The scalar vertex, used in this example,
does not have an infrared divergence. However, the naive unitary chiral
extrapolation cannot remove a non-perturbative (NP) effect. If one has
a handle on 
the direction of partially quenched mass, the leading NP effect
can be isolated \cite{Aoki:2007xm}. Or one can study the response
of the vertex function to the mass to evaluate its sensitivity to the
low scale. The latter leads to an estimate of 7\% systematic error for
the present DWF case.

\begin{figure}
  \begin{center}
   \vspace*{-24pt}
  \epsfig{file=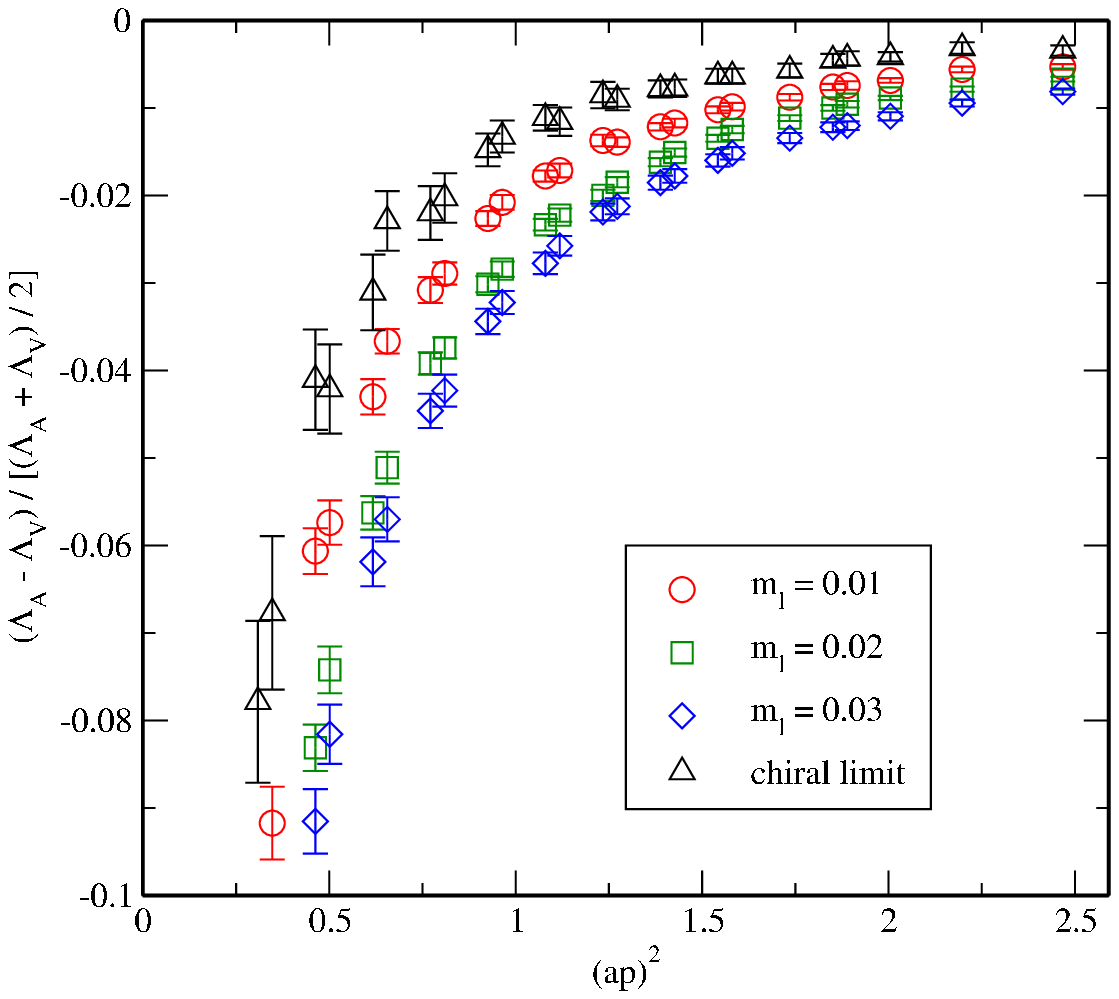, height=7cm}
  \epsfig{file=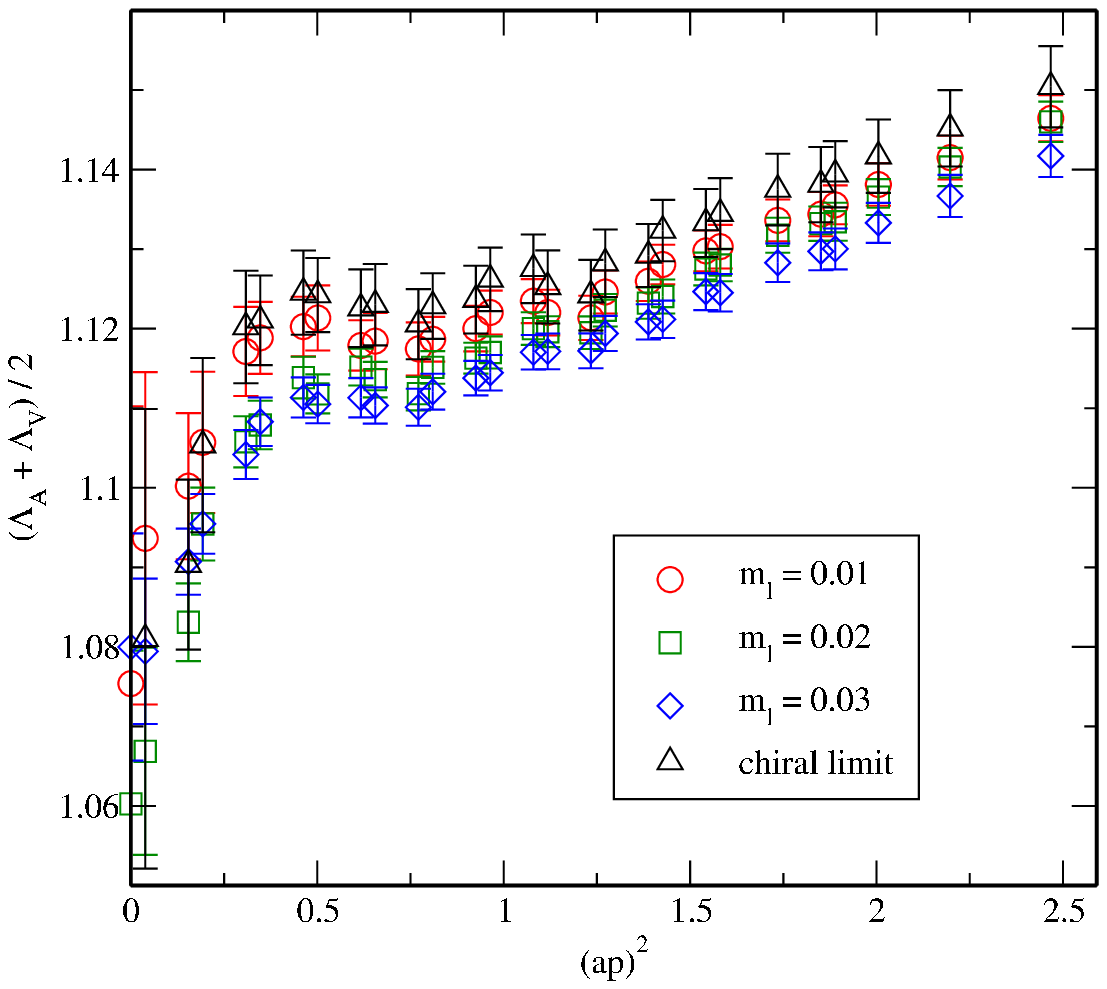, height=7cm}
   \vspace{-18pt}
  \caption{Fractional difference (left panel) and average (right panel)
   of vector and axialvector vertex amplitudes as functions of
   $(pa)^2$ for $n_f=2+1$ DWFs at $a^{-1}\simeq 1.7$ GeV
   \cite{Aoki:2007xm}. Three masses are used to make an extrapolation to
   chiral limit.}
 \label{fig:Lambda_AV}
  \end{center}
\end{figure}

The quark wave function renormalization is calculated from $\Lambda_V$
through Eq.~(\ref{eq:ren_cond_V}).
With a chiral fermion formulation it is often averaged with
the traced axial vector vertex $\Lambda_A$ to gain statistics. In
perturbation theory $\Lambda_V=\Lambda_A$, but this can be violated 
at low momenta. The WT identity, i.e.~the equivalence of
Eqs.~(\ref{eq:ren_cond_V}) and (\ref{eq:RIq}), holds
non-perturbatively for the vector current. As the axial current couples
with the NG boson, the equality is violated non-perturbatively.
Fig.~\ref{fig:Lambda_AV} shows an example of the difference
of $\Lambda_A-\Lambda_V$ as a function of $(pa)^2$. The difference gives
an estimate of the  NP contamination error. We will later use the
window $1.3<(pa)^2<2.5$ where the largest difference is about 1 \% which 
is small compared to the other systematic errors, such as the
aforementioned 7\%. 
The right panel shows the average $(\Lambda_A+\Lambda_V)/2$.
Now the quark mass renormalization is calculated through
\begin{equation}
 Z_m^{\rimom}(\mu) 
  = Z_A \left.\frac{2\Lambda_S}{\Lambda_A+\Lambda_V}\right|_{\mu^2=p^2},
\end{equation}
where $Z_A$ ($=Z_V$) is obtained from a ratio of hadronic two point
correlation functions of partially conserved and local axial vector
current \cite{Aoki:2007xm},
$Z_A=\langle{\mathcal A}_0(x)P(0)\rangle/\langle A_0(x)P(0)\rangle$.
\begin{figure}
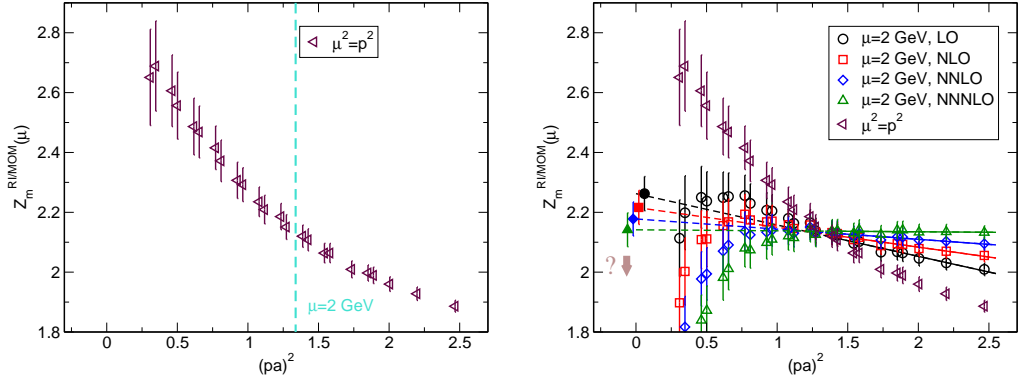

  \begin{center}
  \epsfig{file=Zm_RI_2GeV.eps, height=5cm}
   \hspace{12pt}
  \epsfig{file=Zm_pureRI_LO2NNNLO+.eps, height=5cm}
  \caption{Left panel shows $Z_m^{\rimom}(\mu)$ with
   $\mu^2=(pa)^2$, where cyan line indicates $p^2=(2{\rm GeV})^2$. 
   Right panel shows same data and those with RG running stripped
   off to get $Z_m^{\rimom}(2{\rm GeV})$ by LO to NNNLO PT as a
   function of matching scale.}
 \label{fig:Z_m_RI}
  \end{center}
\end{figure}

The left panel of
Fig.~\ref{fig:Z_m_RI} shows $Z_m^{\rimom}(\mu)$ with $\mu^2=p^2$ 
as a function of $(pa)^2$. $Z_m^{\rimom}(2 {\rm GeV})$, for example,
can be read off from the intercept with the cyan line drawn at 
$p^2=(2{\rm GeV})^2$. This naive determination, however, might suffer from
discretization errors, as well as the NP contamination error which has already
been estimated. 

For the discretization error a common procedure is to
strip off the renormalization-group (RG) running with perturbation
theory. If the truncation 
error is negligible in a certain range of momenta, a linear extrapolation
to $(pa)^2=0$ from the range will remove a leading $(pa)^2$ error. However,
as a comparison between different orders of perturbation theory indicates,
this extrapolation does not always give a better estimate than the
simple method from the intercept (left). 
A particular problem of the extrapolation is that the remnant truncation
error of the perturbative running gets enhanced by $(pa)^2\to 0$.
Thus, if there is any indication of sizable PT truncation error,
as is indicated in the figure, 
$(pa)^2\to 0$ should not been performed.

Fixing the matching scale $p$ to a non-zero value could result in a biased
estimate of the renormalization factor due to a discretization error 
$O(p^2a^2)$ when only one lattice spacing is available.  However,
keeping the scale in physical unit for instance $p=2$ GeV while taking
the continuum limit 
of the renormalized quantity will take away the leading $(pa)^2$
error automatically. This is clearly better than taking the $(pa)^2\to 0$ 
limit at fixed $a$ when a sizable truncation error of PT is suspected.

As the $(pa)^2$ dependence for $Z_m$ in the figure is flat with 
the NNNLO running, it does not matter whether to take the
$(pa)^2\to 0$ or not in this particular case.
The $(pa)^2\to 0$ value was adopted for $Z_m^{\rimom}(2 {\rm GeV})$ in
\cite{Aoki:2007xm}.  
$Z_m^{\msbar}(2 {\rm GeV})$ is obtained by multiplying the conversion
factor calculated to three loops 
\cite{Chetyrkin:1999pq}, Eq.~(\ref{CmRIMOM}). The truncation error
is estimated from the size of the three-loop term, which turns out to be
6\%. The resulting
renormalization factor is $Z_m^{\msbar}(2{\rm GeV})=1.656(157)$,
where all the errors (mass dependence 7\%,
$\Lambda_A-\Lambda_V$ 1\%, PT truncation error 6\%)
have been summed up in quadrature with the statistical error.

\subsection{RI/MOM new developments}
\label{subsec:rimom_dev}

As described above there are three competing effects for possible systematic
uncertainty in the RI/MOM schemes which are:
\begin{enumerate}
 \item unwanted non-perturbative effects,
 \item $(pa)^2$ and higher,
 \item $\alpha_s^{n+1}(\mu)$ and higher.
\end{enumerate}
The last error is relevant for the renormalization factor in
$\MSbar$, or in the case where the perturbative running are used
for whatever purpose.
There are several ideas to overcome one of them or aiming to reduce most
of them. Here the methods are explained for the recent large scale
calculations for the quark mass renormalization by four collaborations:
ETM for $n_f=2$ twisted mass 
fermions with RI'/MOM \cite{Dimopoulos_pc,Constantinou:2010gr},
QCDSF/UKQCD for $n_f=2$ improved Wilson fermions with RI'/MOM
\cite{Gockeler_pc,Gockeler:2010yr}, 
JLQCD/TWQCD for $n_f=2$ and $2+1$ overlap fermions with RI/MOM
\cite{Noaki:2009xi}, and RBC/UKQCD for $n_f=2+1$ DWFs with RI/SMOM
and $\RISMOMgm$ \cite{RBC_UKQCD_in_prep}.

For 1, ETM, QCDSF/UKQCD and JLQCD/TWQCD perform the subtraction.
ETM and JLQCD/TWQCD exploit the partially quenched data for the fit to
subtract the leading effects.
JLQCD/TWQCD also performed the RI/MOM analysis on the $\epsilon$-regime
where the SSB effect is suppressed. They demonstrated that
$\Lambda_A-\Lambda_V=0$ holds well in the region of $p^2$ where 
even a few percent difference was observed with p-regime data.
RBC/UKQCD uses SMOM schemes (see Sec.~\ref{sec:rismom}).

For 2, ETM performs $O(g^2a^2)$ subtraction for terms like
$(pa)^2$ and $\sum_\mu (p^4_\mu/p^2) a^2$, obtained through 
analytic calculation in lattice perturbation theory \cite{Constantinou:2009tr}.
QCDSF/UKQCD uses a similar subtraction, but, with a numerical integration
in lattice perturbation theory. The two collaborations extrapolate
remnant $(pa)^2\to 0$.
RBC/UKQCD fixes $p=2$ GeV, expecting the continuum extrapolation of the
renormalized quantity with $a^2\to 0$ removes the $(pa)^2$ error.
As a second method ETM performs the similar procedure as RBC/UKQCD and
they checked the consistency with the first method after the continuum
extrapolation. 

The RI/MOM to $\MSbar$ conversion factor for the quark mass is available
to $\alpha_s^3$ \cite{Chetyrkin:1999pq}.
JLQCD/TWQCD estimates the size of the $\alpha_s^4$ contribution 
from the extrapolation of $\alpha_s^n$ contribution for $n\le 3$
and take it as the truncation error. RBC/UKQCD study the difference of
$Z_m^\msbar$ 
from two different RI/SMOM intermediate schemes for the truncation error
of $\RISMOM\to\MSbar$ perturbative conversion. They also check the 
consistency of the error with the size of the highest order
in the conversion factor.

\subsection{RI/SMOM schemes}
\label{sec:rismom}

It was demonstrated in Ref.~\cite{Aoki:2007xm} that the severe infrared
sensitivity of the RI/MOM scheme can significantly be reduced  by avoiding
the use of exceptional momenta. The argument stems from Weinberg's
theorem \cite{Weinberg:1959nj} on the behavior of the vertex function
for large external momenta, where a set of external momenta which has
zero partial sum is called exceptional. 

\begin{figure}
  \begin{center}
   \epsfig{file=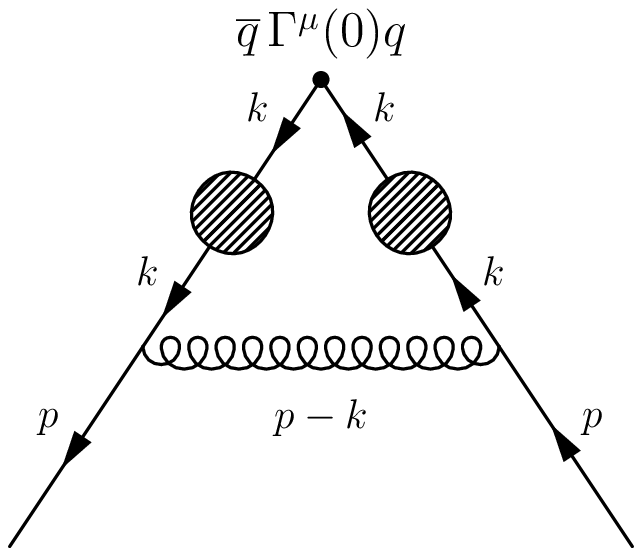, height=3.6cm}
   \hspace{2cm}
   \epsfig{file=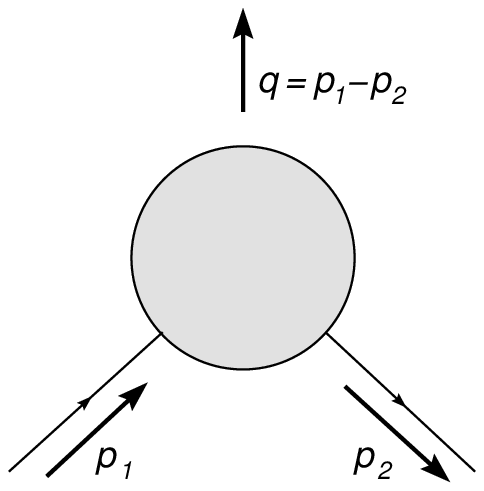, height=3.6cm}
  \caption{A sample figure which makes $1/p^2$ contribution with SSB
   from Ref.~\cite{Aoki:2007xm} (left) and general momentum configuration
   of bilinear vertex (right) including the exceptional ($q=0$) and
   symmetric non-exceptional ($p_1^2=p_2^2=q^2$) momenta from
   Ref.~\cite{Sturm:2009kb}. 
}
 \label{fig:Gamma_bilinear}
  \end{center}
\end{figure}
In the case of bilinear operators with
exceptional momenta, the momentum that flows in from one fermion leg
can be rerouted to the other leg by one-gluon exchange
(see Fig.~\ref{fig:Gamma_bilinear}). If this happens
the operator part allows a small momentum flow, which triggers SSB.
This contamination only suppresses as $1/p^2$ at high momenta, hence
could be sizable for the typical allowed momentum in simulations.
This can be avoided by eliminating the exceptional channel by setting
all $p_1$, $p_2$ and $q$ non-zero in the right panel in
Fig.~\ref{fig:Gamma_bilinear}. 

The $\RISMOM$ and $\RISMOMgm$ schemes are constructed for the
quark bilinear operators in
Ref.~\cite{Sturm:2009kb}, exploiting the non-exceptional momenta and
carefully ensuring the chiral Ward-Takahashi identities are satisfied.
The symmetric configuration $p_1^2=p_2^2=q^2$ is used, hence,
``symmetric'' MOM scheme. 
A trial calculation \cite{Aoki:2009ka} showed
the RI/SMOM scheme is a promising alternative to the conventional RI/MOM
scheme with reduced systematic uncertainties.

The $\RISMOM$ scheme uses a different projector
$\Proj\propto\slash{q}q_\mu/q^2$ compared to  $\Proj\propto\gamma_\mu$ in
Eq.~(\ref{eq:ren_cond_V}).
The resulting quark wavefunction renormalization is equivalent to 
that of the RI'/MOM scheme, i.e.~$Z_q^{\rismom}=Z_q^{\ri'mom}$. Another
scheme named $\RISMOMgm$ uses the same projector as the $\RIMOM$ scheme
$\Proj\propto\gamma_\mu$. It should be remembered, though, due to the
use of the symmetric momentum, the resulting wave function
renormalization is different, i.e.~$Z_q^{\rismomgm}\ne Z_q^{\rimom}$.
Not only the reduction of the unwanted non-perturbative effect, but the
better convergence of the PT for the matching to $\MSbar$ from SMOM
schemes are observed. 
The matching was calculated to one-loop \cite{Sturm:2009kb}.
Evaluating the magnitude at each order in $\alpha_s$
\footnote{$\alpha_s^{\msbar (3)}(2{\rm GeV})=0.2907$ from four loop beta
function has been used.},
\begin{eqnarray}
 C_m(\RISMOM\to\MSb,\mu=2{\rm GeV},n_f=3) & = &
  1 - 0.015 + \cdots, \label{eq:CmSMOM}\\
 C_m(\RISMOMgm\to\MSb,\mu=2{\rm GeV},n_f=3) & = &
  1 - 0.045 + \cdots\,. \label{eq:CmSMOMgm}
\end{eqnarray}
In the RI/MOM and RI'/MOM schemes the conversion factors are
known to three-loop order~\cite{Chetyrkin:1999pq,Gracey:2003yr}: 
\begin{eqnarray}
 C_m(\RIMOM\to\MSb,\mu=2{\rm GeV},n_f=3) & = &
  1 - 0.123 - 0.070 - 0.048 + \cdots,\label{CmRIMOM}\\
 C_m(\RI'MOM\to\MSb,\mu=2{\rm GeV},n_f=3) & = &
  1 - 0.123 - 0.065 - 0.044 + \cdots.
\end{eqnarray}
The convergence of the new SMOM schemes to $\MSb$ is better.
Moreover, it has recently been confirmed that this
trend continues to two loops \cite{Gorbahn:2010bf,Almeida:2010ns},
strongly suggesting  the truncation error of the matching is
significantly small for the SMOM schemes.

The new SMOM schemes are being used for the quark mass renormalization
in $n_f=2+1$ DWFs with two lattice spacings by RBC/UKQCD
collaborations \cite{RBC_UKQCD_in_prep}.
\begin{figure}
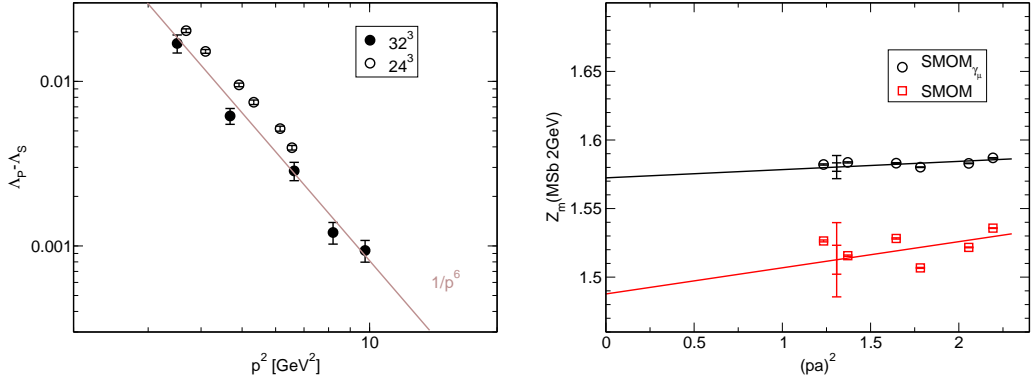

  \begin{center}
   \epsfig{file=Lambda_P-S_scaling-.eps, height=5cm}
   \hspace{12pt}
   \epsfig{file=Zm24.eps, height=5cm}
  \caption{The difference $\Lambda_S-\Lambda_P$ as a function of 
   $p^2$ GeV$^2$ for $a\simeq 0.11$ ($24^3$) and $0.09$ fm ($32^3$) DWF
   lattices (left)    \cite{RBC_UKQCD_in_prep}.
   Solid straight line is proportional $1/p^6$ for guide for eyes.
   Right panel shows $Z_m^{\msbar}(2 {\rm GeV})$ with the SMOM or $\SMOMgm$
   intermediate NPR scheme as function of $(pa)^2$ for the same
   $a\simeq 0.11$ fm DWF lattice.
}
 \label{fig:ZmSMOMs24}
  \end{center}
\end{figure}
The left panel of Fig.~\ref{fig:ZmSMOMs24} shows the difference of
pseudoscalar and scalar vertex amplitude $\Lambda_P-\Lambda_s$,
as an indicator of the unwanted non-perturbative effect,
at the chiral limit. Not only is the difference finite, but
small (1\% at $p=2$ GeV for instance), which is in clear contrast
to the same quantity in the MOM scheme where it was divergent in the
chiral limit as $1/(mp^2)$. Furthermore it exhibits the
$1/p^6$ behaviour expected from Weinberg's power counting and the
possible symmetry breaking pattern for the non-exceptional momenta
\cite{Aoki:2007xm}.

The right panel shows the $Z_m^{\msbar}(2{\rm GeV})$ for the
$a\simeq 0.11$ fm lattice, evaluated with
the $\SMOM$ or $\SMOMgm$ intermediate NPR scheme combined with one-loop
matching \cite{Sturm:2009kb} and two-loop $\MSbar$ running 
as function of the matching scale squared.  The origin of the inconsistency
could be the truncation error of the perturbation theory and the lattice
discretization error.
The non-perturbative effect
is too small to explain this difference.
The existence of the two independent estimates of $Z_m^{\msbar}$ helps
to estimate the systematic error.
A preliminary estimate of the central value from an error weighted average
at $p=2$ GeV, taking into account the spread of the point around the
linear fit, a variation with respect to the change of the matching
scale is given as $Z_m^{\msbar}(2{\rm GeV})=1.568(75)$, where the
error includes an estimate of PT truncation error.
This value is consistent with the same quantity obtained with the
conventional RI/MOM scheme, $Z_m^{\msbar}(2{\rm GeV})=1.656(157)$
\cite{Aoki:2007xm}, but with a considerably reduced error.
Further error reduction is expected by incorporating the two-loop
matching factors.

The non-exceptional momentum configuration has been applied to the
renormalization of non standard-model $B_K$ operators, where
the usual RI/MOM kinematics produces $1/(mp^2)$ type infrared
divergence. The non-exceptional momenta removed these divergences
completely as expected \cite{RBC_UKQCD_in_prep}.

An extension of SMOM kinematics to the standard-model four Fermi
operator for $B_K$ 
uses $\SMOM$ and $\SMOMgm$ $Z_q$.  For the $\Gamma$ projector for
the four quark operator, one can take the conventional type as well
as the one with momentum transfer $q_\mu$ in analogy to the $\SMOM$ 
and $\SMOMgm$ projector. The $2\times 2$ combinations make four
$\SMOM$ schemes. Including the original $\RIMOM$ scheme,
five independent schemes lead to a better estimate of the central
value of $Z_{B_K}^{\rm NDR}(2 {\rm GeV})$ and the systematic error
\cite{RBC_UKQCD_in_prep}.

The advantages and remaining issues of the SMOM schemes
are summarized as follows. Some have already been mentioned. 
The RI/SMOM schemes improved from the RI/MOM scheme on:
\begin{itemize}
 \item Contamination of the unwanted non-perturbative SSB effect is
       greatly reduced.
 \item Mass dependence is very weak.
 \item PT matching to $\MSbar$ has far better convergence.
 \item Signal/noise ratio for the non-exceptional momentum is improved
       from the exceptional.
\end{itemize}
The second point is consistent with the first point and particularly
convenient for $n_f=2+1$ calculations, 
where one can safely keep the strange mass around the physical point.
The last point is likely due to the insensitivity
of non-exceptional vertex functions to the low energy fluctuation.
There could be a room for an improvement on
\begin{itemize}
 \item enhanced effect of $O(4)$ breaking in momentum space.
\end{itemize}
The last issue is visible in the right panel of
Fig.~\ref{fig:ZmSMOMs24} especially for the $\RISMOM$ scheme.
It manifests itself as a non-smooth behaviour in $(pa)^2$. 
Such an effect persists also in the MOM scheme, but smaller. The leading
$O(4)$ breaking term appears at $O(g^2a^2)$ with a structure
of $\sum_\mu p_\mu^4/p^2$ \cite{Constantinou:2009tr} for MOM scheme.
As there is no precise knowledge of such terms in SMOM schemes,
the effect has just been estimated from the spread of the data and
taken into account in the systematic error.
Given that now the two-loop conversion factor is available,
this error could dominate the systematic error in the near future,
then the solution to this issue will be desired.

\subsection{Other remarks in RI/MOM schemes}
\label{sec:rimom_remarks}

For DWFs at finite fifth dimension $L_s$ the axial
Ward-Takahashi identity has an additional contribution to the naive
extension from the continuum one from the pseudoscalar like the
operator at mid point 
of fifth dimension. As such the partially conserved axial current
${\mathcal A}^a_\mu$
does not conserve in the chiral limit, which makes the renormalization
factor of the current deviate from 1 \cite{Sharpe:2007yd}. 
Denoting the additive renormalization of the quark mass in lattice units due to 
the non-vanishing midpoint term as $\mres$, 
the effect was estimated as $Z_{\mathcal A}=1+O(\mres)$ for the perturbative
and $Z_{\mathcal A}=1+O(\mres^2)$ for the non-perturbative contribution
\cite{Allton:2008pn}. The size of the correction was estimated to be no
larger than 1\% for $n_f=2+1$ DWFs at $a\simeq 0.11$ fm.
As the ratio of the local and partially conserved axial current 
is an input to the RI/MOM schemes to evaluate the quark
wavefunction renormalization, this systematic error will propagate to
the renormalization factors for multi-quark operators and the quark mass.
This is not a dominant error in the current calculations.
However, if the other errors are getting smaller, eventually this issue will
need to be revisited.

It is reported in Ref.~\cite{Aoki:2007xm} that the use of point sources
for the $\RIMOM$ scheme can result in underestimating
the statistical error. The problem is enhanced towards the larger momenta.
This is understandable because a larger momentum in 
the Fourier transformation of the sink position tends to sample a narrower region 
in the real space, thus effectively reduces the statistics.
One solution is to have several distant source positions or to have
a random source position. Another is the use of a volume momentum source, which 
effectively samples the entire volume.

\section{Developments in Schr\"odinger functional scheme}

Finite volume scheme with Schr\"odinger functional (SF) boundary conditions
combined with a step 
scaling analysis provides a solid framework to perform a precision
non-perturbative renormalization \cite{Luscher:1992an}.
The renormalization group running of the operators in the continuum
theory can be calculated in a very efficient manner.
Possible applications include the determination of
the running coupling constant, renormalization of multi-quark operators,
coefficients of Symanzik improvement operators.
The operator renormalization condition is applied to correlation functions
of the operator in the bulk and distant boundary field(s). As such,
the scheme is compatible for the Symanzik on-shell improvement of
the operators.
The practical difficulty is that the scheme needs tailored gauge
ensemble generation for many parameter points for the step scaling
and continuum limit. However, once the running of the particular operator
is determined (with any gauge-fermion action), the remaining process
is to perform the matching at given lattice spacing.

The technique has been explained to some detail in the plenary presentations
in the past conferences (see \cite{Sint:2000vc,Sommer:2002en}).
In the plenary talk of Lattice 2002 by Sommer, who gave the last 
plenary presentation dedicated on the non-perturbative renormalization
until this year, he listed three 
problems which remain to be tackled. All related to the SF
scheme: 
\begin{itemize}
 \item Development of SF scheme for HQET at $1/m_h$,
 \item $n_f=3$ simulation for non-perturbative estimate of
       $\alpha_{\rm SF}(\mu)$,
 \item applicability of SF scheme for the chiral fermions.
\end{itemize}
By this year the answers of these questions have been given.
The non-perturbative renormalization of HQET formulation at $1/m_h$
is important to b-quark physics. 
In a recent lattice conference the calculation strategy has been
explained in detail by Della Morte \cite{DellaMorte:2007ny}.
This year the Alpha collaboration reported on the calculation of ground state
and first excited state spectrum of bottom-strange mesons, as well
as the decay constant of $B_s$, with completing the continuum
extrapolation \cite{Blossier:2009mg}.
Application to $n_f=2$ improved Wilson fermion is underway.

$B^0-\overline{B^0}$ mixing matrix elements have great importance
to flavor physics (for example to 
constrain the apex of the unitarity triangle of CKM matrix
\cite{Aubin:2009yh,VandeWater:2009uc}).
So far the renormalized operators are constructed in the static
approximation to b-quark for $n_f=2$ \cite{Palombi:2007dr} and
$n_f=2$ \cite{Dimopoulos:2007ht}. 
If the $O(1/m_h)$ operator is constructed, the systematic error
from $O(1/m_h^2)$ could be constrained to sub-percent level.
However, it is yet to be realized.

CP-PACS and PACS-CS collaborations have been performing $n_f=2+1$
simulations with non-perturbatively improved Wilson fermions.
SF coupling $\alpha_{\rm SF}$ is obtained by step scaling to give
a high accuracy estimate of $n_f=3$ coupling at a high scale,
where matching to $\MSbar$ scheme is performed. Running down to
charm threshold $\mu=m_c$ by perturbative running to mach to $n_f=4$
and then running up to $n_f=5$, $\mu=m_Z$, they obtained
$\alpha_s(m_Z)$ \cite{Aoki:2009tf}. See Sec.~\ref{sec:alpha_s} for
comparison with the other estimate.
They push the project further to calculate the quark mass
renormalization through SF scheme. The preliminary results
are reported in \cite{Taniguchi:2009hb}. 
The resulting renormalized quark mass is discussed in Sec.~\ref{sec:mass}
together with those from other discretization/NPR schemes.

The SF boundary condition on the quark propagator naively interferes
with the Ginsparg-Wilson (GW) relation at the boundary. As such,
the GW relation has to be modified to properly define the SF formalism
on overlap or domain-wall fermions (DWFs). The solution has been
provided by Taniguchi for quenched overlap \cite{Taniguchi:2004gf} 
and DW \cite{Taniguchi:2006qw} fermions through orbifolding.
This method is successfully applied to the renormalization for the light quark
mass and $B_K$ with DWFs in the quenched approximation \cite{Nakamura:2008xz}.
It is extended by Sint \cite{Sint:2005qz} for the even number of flavors case,
and L\"uscher \cite{Luscher:2006df} for any number of flavors for overlap
fermions. This year further development has been reported by Takeda
\cite{Takeda:2009sk} on SF scheme construction in DWF with any number of
fermions. 

The SF scheme is being used for twisted mass simulations as well.
ETM collaboration tune the boundary counter term of ``chirally rotated
SF'' for twisted mass fermions non-perturbatively \cite{Lopez:2009yc} to
ensure the automatic $O(a)$ improvement.
The tuning has been performed for the quenched approximation. Application to
$n_f=2$ is underway.

\section{Application specific developments/remarks}

In this section application specific developments in recent years
and some comparison of different schemes are provided.

\subsection{Strong coupling constant $\alpha_s$}
\label{sec:alpha_s}

Due to the availability of several independent $n_f=2+1$ simulations in recent
years, lattice estimates of $\alpha_s^{\msbar} (M_Z)$ have now become
very strong 
compared to the other methods. This has not been possible when
only quench or $n_f=2$ simulations were available, because of the
inaccessibility of the threshold scale $\mu=m_{u,d}$ or $m_s$ in perturbation
theory. An estimate of $\alpha_s(\mu)$ at $n_f=3$ for $\mu=O(1)$ GeV allows
one to use perturbative matching at charm threshold $\mu=m_c$ by 
perturbation theory (PT) to
$n_f=4$, and then at $\mu=m_b$ to $n_f=5$, with RG evolution with 
anomalous dimension calculated with PT. The systematic error which
arises in this calculation apart from the lattice non-perturbative part
mainly comes from 1) the PT machining of a particular scheme used in
NP part to $\MSbar$ and 2) the $n_f=3\leftrightarrow 4$ matching at
$\mu=m_c$ in $\MSbar$ scheme.  Due to 2), one needs to select a scheme
with which the matching is available to high order typically to
$\alpha_s^3$. 

\begin{figure}
  \begin{center}
  \epsfig{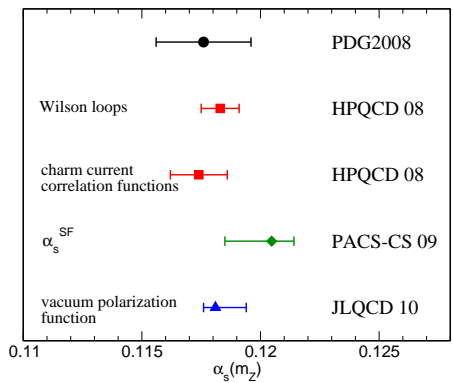}
  \caption{Summary of $\alpha_s^{\msbar}(M_Z)$ from $n_f=2+1$
   lattice simulations, compared with PDG 2008 average
   \cite{Amsler:2008zzb} (black).
   Results are from Wilson loop \cite{Davies:2008sw} and charm
   correlation functions \cite{Allison:2008xk} with staggered (red),
   SF coupling \cite{Aoki:2009tf} with Wilson (green),
   and vacuum polarization function \cite{Shintani:2010ph} with overlap
   fermions (blue). 
   }
 \label{fig:alpha_s}
  \end{center}
\end{figure}
There have been four estimates of $\alpha_s^{\msbar}(M_Z)$ by the
conference, 
1) through $\alpha_V$ calculated from Wilson loops
  \cite{Mason:2005zx,Davies:2008sw}, 
2) from moments of the charmonium current correlators \cite{Allison:2008xk}, 
3) from SF coupling \cite{Aoki:2009tf},
4) from vacuum polarization function
  \cite{Shintani:pos_lat09,Shintani:2010ph}. 
For 1) one calculates Wilson loops to get $\alpha_V(\mu)$ through
lattice perturbation theory, and then matched to
$\alpha_s^{\msbar}(\mu)$ at three-loop (NNNLO).
2) uses the continuum three-loop expression (NNNLO) of a moment of charm
current-current correlation function.
This is a by-product of their charm quark mass estimate for which they use
same correlation functions.
3) use step scaling of the SF coupling and matching to $\MSbar$.
Although this has been obtained only to two loops (NNLO), the matching is
performed at high scale $\mu\gg m_c$, where the PT truncation error is
small compared to the other errors. The renormalization scale is
set from the inverse linear lattice extension $\mu=1/L$. Once
$\alpha_s^{\msbar(3)}(\mu)$ 
has been obtained, it is run down to $\mu'=m_c$ to match to $n_f=4$.
This matching and running are done with NNNLO.
4) the continuum vacuum polarization function has been obtained through
operator product expansion and the relevant coefficients has been
calculated to NNNLO. The renormalization scale is set from the size
of the injected momentum at the current.
A compiled summary of $\alpha_s^{\msbar}(M_Z)$ estimates is shown in
Fig.~\ref{fig:alpha_s}. Results from different method / lattice discretization 
are consistent with each other and with the PDG2008
average \cite{Amsler:2008zzb}.

All of these calculations reached a remarkable precision of the strong
coupling constant. One has to keep in mind, however, all of them uses
the perturbation theory at $\mu\simeq m_c$, where non-perturbative effect
might still be important. Analysis on $n_f=2+1+1$ simulations on
finer lattices will be able to address this issue.

There are several technical development on determining $\Lambda_\msbar$
using the Landau-gauge ghost and gluon propagators,
which has been applied to $n_f=0$ and $2$
\cite{Boucaud:2008gn,Desoto:2009aw,Sternbeck:2010xu}.

\subsection{Light quark mass}
\label{sec:mass}

\begin{figure}
  \begin{center}
  \epsfig{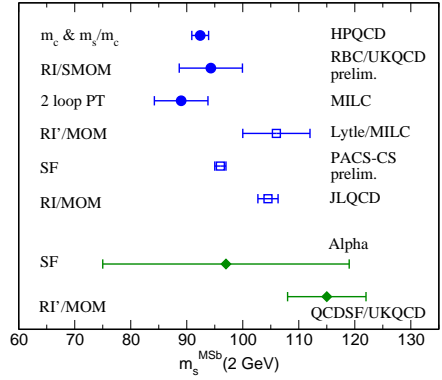}
  \caption{Summary of strange mass $m_s^\msbar(2 {\rm GeV})$
   for $n_f=2$ and $2+1$ simulations. Blue solid symbols
   are from $n_f=2+1$ simulations with continuum extrapolation
   \cite{Davies:2009ih,Mawhinney:2009jy,Bazavov:2009tw}.
   Blue open symbols are from $n_f=2+1$ simulations at single non-zero lattice
   spacing \cite{Lytle:2009xm,Noaki:2009sk,Aoki:2009ix}.
   Green symbols are from $n_f=2$ calculations \cite{DellaMorte:2005kg}
   \cite{Gockeler:2006vi}.
   }
 \label{fig:m_s}
  \end{center}
\end{figure}
A comparison of the light quark mass from different lattice
discretization and renormalization schemes is presented by Scholz
\cite{Scholz:2009yz} in this conference.  Here for the purpose of
discussing the renormalization issues the strange mass compilation is
made with the available data in the literature (Fig.~\ref{fig:m_s}).

For $n_f=2+1$ dynamical fermion computations,
a remarkable precision has been achieved by HPQCD collaboration
\cite{McNeile:2009eq,Davies:2009ih} who exploits
a precise estimate of $\MSbar$ charm quark mass from the
charmonium current correlator and the bare mass ratio $m_s/m_c$.
The use of highly improved staggered quark (HISQ) and the fine lattice
spacing made it possible to use the same action for both light
and charm quarks. This method does not require the conventional lattice
renormalization procedure. 
RBC and UKQCD collaborations \cite{Mawhinney:2009jy} have used RI/SMOM
schemes described in Sec.~\ref{sec:rismom} on DWFs with two lattice
spacings for the continuum extrapolation.
The MILC collaboration \cite{Bazavov:2009tw} uses Asqtad action with two loop
lattice perturbation theory \cite{Mason:2005bj} for the renormalization.
These three results in the continuum limit agree with each other.
One cannot find any issue for the perturbative vs.~non-perturbative
renormalization here.

For results from single lattice spacing, Lytle reported the strange
mass with MILC coarse lattice with RI'/MOM NPR \cite{Lytle:2009xm}.
He compared with the strange mass with perturbative renormalization on the
same lattice and found NPR gives about 20\% larger value.
PACS-CS reported the calculation from improved Wilson fermions directly
simulated on the physical $ud$ and $s$ point with Schr\"odinger
functional NPR \cite{Taniguchi:2009hb,Aoki:2009ix}, 
which has become 30\% larger from their 
earlier estimate with perturbative renormalization referred in Scholz's
compilation \cite{Scholz:2009yz}.
These differences between the perturbative vs.~non-perturbative
renormalization are not necessarily inconsistent with the observed
consistency for the values after the continuum extrapolation.
This is because the difference could arise from the discretization error
in different renormalization schemes.
JLQCD/TWQCD has the estimate from overlap fermions with RI/MOM NPR.
For these three studies, extension to the second lattice spacing are
anticipated. 

There has been several $n_f=2$ calculations reported for the light
quark mass. In this figure of $m_s$ compilation only earlier studies
with Wilson fermions by 
Alpha \cite{DellaMorte:2005kg} and QCDSF-UKQCD \cite{Gockeler:2006vi}
collaboration are shown. The new calculation of average $ud$ quark mass
in the continuum limit with twisted mass fermions combined with the
RI'/MOM scheme has been reported 
\cite{Dimopoulos:2009hy,Baron:2009wt}. The result is consistent with
$n_f=2+1$ continuum values \cite{Scholz:2009yz}.

\subsection{Three quark operators}

A class of three quark operators which are a part of baryon-number
violating four-fermion operators induced by OPE from grand unified
theories are important for the estimate of proton lifetime.
The operator classification,
construction of RI/MOM scheme, and one-loop matching to $\MSbar$ have been
carried out in Ref.~\cite{Aoki:2006ib}. 

The RI/MOM scheme of more general three-quark operators are
constructed by QCDSF-UKQCD \cite{Gockeler:2008we},
for the proton decay matrix elements and nucleon distribution amplitude
\cite{Braun:2008ur}.

\subsection{Static heavy quark}

As mentioned, the SF scheme has been developed for b-quark system
namely heavy-light bilinear and $\Delta F=2$ four quark operators
with the static approximation.
Application of the RI/MOM scheme to this system was tried 
long ago \cite{Donini:1995zd}. The subtlety to have 
a mass independent scheme for the static quark formulation,
where the self-energy has power divergence, prevented further
development. In this conference, Izubuchi showed his study on
the static quark self-energy, hoping to resolve this issue
by making use of the odd part in energy of the Fourier transformed static
quark propagator. Further investigation/demonstration is desired.

\subsection{Relativistic heavy quark}

The Fermilab-Tsukuba approach \cite{ElKhadra:1996mp,Aoki:2001ra}
the heavy quark uses
relativistic clover type fermions. The $(m_h a)^n$ error with all order
in $n$ is removed using perturbation theory. The Columbia group refined the
strategy of this approach to correctly count the independent operator
coefficients \cite{Christ:2006us} and developed the non-perturbative
determination with step-scaling \cite{Li:2008kb}. Later another 
tuning procedure was proposed with by-passing the step scaling
and matching experimental spectrum directly. It has been 
applied to charm \cite{Li:2007en} and bottom \cite{Li:2008kb} systems.

\section{Related development in lattice perturbation theory}

Lattice perturbation theory has been developed for some
quantities. For the quark mass renormalization, two-loop results
are available for staggered \cite{Mason:2005bj} and Wilson(W)/clover(C)
fermions \cite{Skouroupathis:2010mq}.
The latter calculates the renormalization constant in RI'/MOM scheme.
Once that is obtained the continuum matching can be used to convert
it, for example, to $\MSbar$, where the conversion factor has already been
calculated up to three loops \cite{Chetyrkin:1999pq,Gracey:2003yr}.

$O(g^2a^2)$ calculations have been performed on the quark self-energy
and bilinear operators for W/C \cite{Constantinou:2009tr},
twist two bilinears for W/C/twisted mass (tm) 
\cite{Constantinou:2010dn}, four 
Fermi operators for W/tm fermions \cite{Constantinou:2010wn}.
The first one has been used in combination with the 
non-perturbative RI'/MOM renormalization to have $O(g^2a^2)$ effect on
$p^2$ and $\sum_\mu p_\mu^4/p^2$ subtracted (see Sec.~\ref{subsec:rimom_dev}).

Numerical stochastic perturbation theory (NSPS) \cite{DiRenzo:2004ge}
is being developed for the improved gauge and Wilson fermion action
on the quark self-energy \cite{Brambilla:2010qb} at three loop
through the RI'/MOM renormalization condition imposed in the
lattice perturbation theory. Two-loop result for the
Symanzik improvement coefficient $c_{\rm SW}$ and
$c_A$ for Wilson fermions has been obtained \cite{Torrero:2009kc}.
The Landau-gauge ghost propagator has been calculated to three loop
\cite{DiRenzo:2009ni}, anticipating the application to estimate of $\alpha_s$.

\section{Concluding remarks}

Today, the $n_f=2+1$ simulations, incorporating the full dynamics
of three lighter quarks that  cannot be accessed from the perturbation
theory,  are available with four different fermion discretizations.
Thus various crosschecks on the calculated quantities can be
performed. Furthermore, $n_f=2+1+1$ simulations, incorporating the
dynamical charm effect, for two fermion actions are being carried out
\cite{Baron:2009zq,Bazavov:2009wm}.
In these circumstances the renormalization in lattice QCD has become more
important than ever before. Non-perturbative renormalization, which was
reviewed here is a powerful tool to tackle the precision calculation of the 
fundamental parameters of standard model and hadron-operator matrix
elements. In recent years, the Schr\"odinger functional scheme has been applied
to the strong coupling constant and light quark masses for $n_f=3$, 
non-perturbative tuning of the parameters for the simulations,
as well as the b-quark systems.
Due to the window problem, RI/MOM schemes at the typical lattice spacings
of current simulations could suffer from larger systematic errors
than the SF scheme.  Yet, there has been several ideas of bringing
the error under control, which have been discussed in this review.
As a promising new direction the RI/SMOM
schemes were described in some detail. A small summary was made for the
development on the lattice perturbation theory, as well.
The comparison of the strong coupling constant and the strange
quark mass estimate in the continuum limit have shown consistency
despite of the difference of the lattice action and renormalization method used.
The success stands upon the cultivation of not only one but
multiple relevant technologies in the lattice QCD community over the
last two decades, and would motivate further improvement/development
in this stimulating field.

\section*{Acknowledgments}
I would like to thank the organizers of Lattice 2009
for the invitation of this presentation.
I would like to thank the correspondence and discussion for the
material presented here with
Petros Dimopoulos,
Meinulf G\"ockeler,
Jenifer Gonzalez Lopez,
Tereza Mendes,
Jun Noaki,
Haris Panagopoulos,
Olivier Pene,
Rainer Sommer,
Shinji Takeda,
Yusuke Taniguchi,
and Hartmut Wittig.
I am indebted for the presented results from RBC and UKQCD
collaborations to 
Peter Boyle,
Dirk Br\"ommel,
Norman Christ,
Chris Dawson,
Taku Izubuchi,
Chris Kelly,
Chris Sachrajda,
Enno Scholz,
Amarjit Soni,
Christian Sturm,
Jan Wennekers
and other colleagues in the collaborations.
I would like to thank Dirk Br\"ommel
for the critical reading and
comment on the manuscript. 
It is a great regret that my close collaborator
Jan Wennekers passed away in December 2009.
This work is supported in part by JSPS Kakenhi Grant No.~21540289.

\bibliography{all}

\providecommand{\href}[2]{#2}\begingroup\raggedright\begin{thebibliography}{10}

\bibitem{Rossi:1996yu}
G.~C. Rossi, {\it {Non-perturbative renormalization in lattice QCD}},  {\em
  Nucl. Phys. Proc. Suppl.} {\bf 53} (1997) 3--15,
  [\href{http://xxx.lanl.gov/abs/hep-lat/9609038}{{\tt hep-lat/9609038}}].

\bibitem{Testa:1997ne}
M.~Testa, {\it {Non-perturbative renormalisation and kaon physics}},  {\em
  Nucl. Phys. Proc. Suppl.} {\bf 63} (1998) 38--46,
  [\href{http://xxx.lanl.gov/abs/hep-lat/9709044}{{\tt hep-lat/9709044}}].

\bibitem{Sint:2000vc}
S.~Sint, {\it {Non-perturbative renormalization in lattice field theory}},
  {\em Nucl. Phys. Proc. Suppl.} {\bf 94} (2001) 79--94,
  [\href{http://xxx.lanl.gov/abs/hep-lat/0011081}{{\tt hep-lat/0011081}}].

\bibitem{Sommer:2002en}
R.~Sommer, {\it Non-perturbative renormalization of hqet and qcd},  {\em Nucl.
  Phys. Proc. Suppl.} {\bf 119} (2003) 185--197,
  [\href{http://xxx.lanl.gov/abs/hep-lat/0209162}{{\tt hep-lat/0209162}}].

\bibitem{DellaMorte:2007ny}
M.~Della~Morte, {\it {Standard Model parameters and heavy quarks on the
  lattice}},  {\em PoS} {\bf LAT2007} (2007) 008,
  [\href{http://xxx.lanl.gov/abs/0711.3160}{{\tt arXiv:0711.3160}}].

\bibitem{Martinelli:1993dq}
G.~Martinelli, S.~Petrarca, C.~T. Sachrajda, and A.~Vladikas, {\it
  {Nonperturbative renormalization of two quark operators with an improved
  lattice fermion action}},  {\em Phys. Lett.} {\bf B311} (1993) 241--248.

\bibitem{Martinelli:1994ty}
G.~Martinelli, C.~Pittori, C.~T. Sachrajda, M.~Testa, and A.~Vladikas, {\it {A
  General method for nonperturbative renormalization of lattice operators}},
  {\em Nucl. Phys.} {\bf B445} (1995) 81--108,
  [\href{http://xxx.lanl.gov/abs/hep-lat/9411010}{{\tt hep-lat/9411010}}].

\bibitem{Chetyrkin:1999pq}
K.~G. Chetyrkin and A.~Retey, {\it Renormalization and running of quark mass
  and field in the regularization invariant and ms-bar schemes at three and
  four loops},  {\em Nucl. Phys.} {\bf B583} (2000) 3--34,
  [\href{http://xxx.lanl.gov/abs/hep-ph/9910332}{{\tt hep-ph/9910332}}].

\bibitem{Aoki:2007xm}
Y.~Aoki {\em et.~al.}, {\it {Non-perturbative renormalization of quark bilinear
  operators and $B_K$ using domain wall fermions}},  {\em Phys. Rev.} {\bf D78}
  (2008) 054510, [\href{http://xxx.lanl.gov/abs/0712.1061}{{\tt
  arXiv:0712.1061}}].

\bibitem{Aoki:2009ka}
{\bf RBC-UKQCD} Collaboration, Y.~Aoki, {\it {Quark mass renormalization with
  non-exceptional momenta}},  {\em PoS} {\bf LATTICE2008} (2008) 222,
  [\href{http://xxx.lanl.gov/abs/0901.2595}{{\tt arXiv:0901.2595}}].

\bibitem{Blum:2001sr}
T.~Blum {\em et.~al.}, {\it {Non-perturbative renormalisation of domain wall
  fermions: Quark bilinears}},  {\em Phys. Rev.} {\bf D66} (2002) 014504,
  [\href{http://xxx.lanl.gov/abs/hep-lat/0102005}{{\tt hep-lat/0102005}}].

\bibitem{Cudell:1998ic}
J.-R. Cudell, A.~Le~Yaouanc, and C.~Pittori, {\it {Pseudoscalar vertex,
  Goldstone boson and quark masses on the lattice}},  {\em Phys. Lett.} {\bf
  B454} (1999) 105--114, [\href{http://xxx.lanl.gov/abs/hep-lat/9810058}{{\tt
  hep-lat/9810058}}].

\bibitem{Giusti:2000jr}
L.~Giusti and A.~Vladikas, {\it {RI/MOM renormalization window and Goldstone
  pole contamination}},  {\em Phys. Lett.} {\bf B488} (2000) 303--312,
  [\href{http://xxx.lanl.gov/abs/hep-lat/0005026}{{\tt hep-lat/0005026}}].

\bibitem{Dimopoulos_pc}
{\bf ETM} Collaboration, private commumication~with P.~Dimopoulos.

\bibitem{Constantinou:2010gr}
M.~Constantinou {\em et.~al.}, {\it {Non-perturbative renormalization of quark
  bilinear operators with Nf=2 (tmQCD) Wilson fermions and the tree- level
  improved gauge action}},  \href{http://xxx.lanl.gov/abs/1004.1115}{{\tt
  arXiv:1004.1115}}.

\bibitem{Gockeler_pc}
{\bf QCDSF-UKQCD} Collaboration, private commumication~with M.~Gockeler.

\bibitem{Gockeler:2010yr}
M.~Gockeler {\em et.~al.}, {\it {Perturbative and Nonperturbative
  Renormalization in Lattice QCD}},
  \href{http://xxx.lanl.gov/abs/1003.5756}{{\tt arXiv:1003.5756}}.

\bibitem{Noaki:2009xi}
J.~Noaki {\em et.~al.}, {\it {Non-perturbative renormalization of bilinear
  operators with dynamical overlap fermions}},  {\em Phys. Rev.} {\bf D81}
  (2010) 034502, [\href{http://xxx.lanl.gov/abs/0907.2751}{{\tt
  arXiv:0907.2751}}].

\bibitem{RBC_UKQCD_in_prep}
{\bf RBC-UKQCD} Collaboration, in~preparation.

\bibitem{Constantinou:2009tr}
M.~Constantinou, V.~Lubicz, H.~Panagopoulos, and F.~Stylianou, {\it {$O(a^2)$
  corrections to the one-loop propagator and bilinears of clover fermions with
  Symanzik improved gluons}},  {\em JHEP} {\bf 10} (2009) 064,
  [\href{http://xxx.lanl.gov/abs/0907.0381}{{\tt arXiv:0907.0381}}].

\bibitem{Weinberg:1959nj}
S.~Weinberg, {\it High-energy behavior in quantum field theory},  {\em Phys.
  Rev.} {\bf 118} (1960) 838--849.

\bibitem{Sturm:2009kb}
C.~Sturm {\em et.~al.}, {\it {Renormalization of quark bilinear operators in a
  momentum- subtraction scheme with a nonexceptional subtraction point}},  {\em
  Phys. Rev.} {\bf D80} (2009) 014501,
  [\href{http://xxx.lanl.gov/abs/0901.2599}{{\tt arXiv:0901.2599}}].

\bibitem{Gracey:2003yr}
J.~A. Gracey, {\it Three loop anomalous dimension of non-singlet quark currents
  in the ri' scheme},  {\em Nucl. Phys.} {\bf B662} (2003) 247--278,
  [\href{http://xxx.lanl.gov/abs/hep-ph/0304113}{{\tt hep-ph/0304113}}].

\bibitem{Gorbahn:2010bf}
M.~Gorbahn and S.~Jager, {\it {Precise MS-bar light-quark masses from lattice
  QCD in the RI/SMOM scheme}},  \href{http://xxx.lanl.gov/abs/1004.3997}{{\tt
  arXiv:1004.3997}}.

\bibitem{Almeida:2010ns}
L.~G. Almeida and C.~Sturm, {\it {Two-loop matching factors for light quark
  masses and three-loop mass anomalous dimensions in the RI/SMOM schemes}},
  \href{http://xxx.lanl.gov/abs/1004.4613}{{\tt arXiv:1004.4613}}.

\bibitem{Sharpe:2007yd}
S.~R. Sharpe, {\it {Future of Chiral Extrapolations with Domain Wall
  Fermions}},  \href{http://xxx.lanl.gov/abs/0706.0218}{{\tt 0706.0218}}.

\bibitem{Allton:2008pn}
{\bf RBC-UKQCD} Collaboration, C.~Allton {\em et.~al.}, {\it {Physical Results
  from 2+1 Flavor Domain Wall QCD and SU(2) Chiral Perturbation Theory}},  {\em
  Phys. Rev.} {\bf D78} (2008) 114509,
  [\href{http://xxx.lanl.gov/abs/0804.0473}{{\tt arXiv:0804.0473}}].

\bibitem{Luscher:1992an}
M.~Luscher, R.~Narayanan, P.~Weisz, and U.~Wolff, {\it {The Schrodinger
  functional: A Renormalizable probe for nonAbelian gauge theories}},  {\em
  Nucl. Phys.} {\bf B384} (1992) 168--228,
  [\href{http://xxx.lanl.gov/abs/hep-lat/9207009}{{\tt hep-lat/9207009}}].

\bibitem{Blossier:2009mg}
B.~Blossier {\em et.~al.}, {\it {Spectroscopy and Decay Constants from
  Nonperturbative HQET at Order 1/m}},
  \href{http://xxx.lanl.gov/abs/0911.1568}{{\tt arXiv:0911.1568}}.

\bibitem{Aubin:2009yh}
C.~Aubin, {\it {Lattice studies of hadrons with heavy flavors}},
  \href{http://xxx.lanl.gov/abs/0909.2686}{{\tt arXiv:0909.2686}}.

\bibitem{VandeWater:2009uc}
R.~S. Van~de Water, {\it {The CKM matrix and flavor physics from lattice QCD}},
   \href{http://xxx.lanl.gov/abs/0911.3127}{{\tt arXiv:0911.3127}}.

\bibitem{Palombi:2007dr}
F.~Palombi, M.~Papinutto, C.~Pena, and H.~Wittig, {\it {Non-perturbative
  renormalization of static-light four- fermion operators in quenched lattice
  QCD}},  {\em JHEP} {\bf 09} (2007) 062,
  [\href{http://xxx.lanl.gov/abs/0706.4153}{{\tt arXiv:0706.4153}}].

\bibitem{Dimopoulos:2007ht}
{\bf ALPHA} Collaboration, P.~Dimopoulos {\em et.~al.}, {\it {Non-perturbative
  renormalisation of Delta F=2 four-fermion operators in two-flavour QCD}},
  {\em JHEP} {\bf 05} (2008) 065,
  [\href{http://xxx.lanl.gov/abs/0712.2429}{{\tt arXiv:0712.2429}}].

\bibitem{Aoki:2009tf}
{\bf PACS-CS} Collaboration, S.~Aoki {\em et.~al.}, {\it {Precise determination
  of the strong coupling constant in Nf=2+1 lattice QCD with the Schr\'odinger
  functional scheme}},  {\em JHEP} {\bf 10} (2009) 053,
  [\href{http://xxx.lanl.gov/abs/0906.3906}{{\tt arXiv:0906.3906}}].

\bibitem{Taniguchi:2009hb}
{\bf PACS-CS} Collaboration, Y.~Taniguchi, {\it {Determination of the running
  coupling constant $\alpha_s$ for Nf=2+1 QCD with the Schroedinger functional
  scheme}},  {\em PoS} {\bf LAT2009} (2009) 208,
  [\href{http://xxx.lanl.gov/abs/0910.5105}{{\tt arXiv:0910.5105}}].

\bibitem{Taniguchi:2004gf}
Y.~Taniguchi, {\it {Schroedinger functional formalism with Ginsparg-Wilson
  fermion}},  {\em JHEP} {\bf 12} (2005) 037,
  [\href{http://xxx.lanl.gov/abs/hep-lat/0412024}{{\tt hep-lat/0412024}}].

\bibitem{Taniguchi:2006qw}
Y.~Taniguchi, {\it {Schroedinger functional formalism with domain-wall
  fermion}},  {\em JHEP} {\bf 10} (2006) 027,
  [\href{http://xxx.lanl.gov/abs/hep-lat/0604002}{{\tt hep-lat/0604002}}].

\bibitem{Nakamura:2008xz}
{\bf CP-PACS} Collaboration, Y.~Nakamura, S.~Aoki, Y.~Taniguchi, and T.~Yoshie,
  {\it {Precise determination of $B_K$ and right quark masses in quenched
  domain-wall QCD}},  {\em Phys. Rev.} {\bf D78} (2008) 034502,
  [\href{http://xxx.lanl.gov/abs/0803.2569}{{\tt arXiv:0803.2569}}].

\bibitem{Sint:2005qz}
S.~Sint, {\it {The Schroedinger functional with chirally rotated boundary
  conditions}},  {\em PoS} {\bf LAT2005} (2006) 235,
  [\href{http://xxx.lanl.gov/abs/hep-lat/0511034}{{\tt hep-lat/0511034}}].

\bibitem{Luscher:2006df}
M.~Luscher, {\it {The Schroedinger functional in lattice QCD with exact chiral
  symmetry}},  {\em JHEP} {\bf 05} (2006) 042,
  [\href{http://xxx.lanl.gov/abs/hep-lat/0603029}{{\tt hep-lat/0603029}}].

\bibitem{Takeda:2009sk}
S.~Takeda, {\it {A formulation of domain wall fermions in the Schroedinger
  functional}},  \href{http://xxx.lanl.gov/abs/0910.2485}{{\tt
  arXiv:0910.2485}}.

\bibitem{Lopez:2009yc}
J.~G. Lopez, K.~Jansen, D.~B. Renner, and A.~Shindler, {\it {Chirally rotated
  Schroedinger functional: non-perturbative tuning in the quenched
  approximation}},  {\em PoS} {\bf LAT2009} (2009) 199,
  [\href{http://xxx.lanl.gov/abs/0910.3760}{{\tt arXiv:0910.3760}}].

\bibitem{Amsler:2008zzb}
{\bf Particle Data Group} Collaboration, C.~Amsler {\em et.~al.}, {\it {Review
  of particle physics}},  {\em Phys. Lett.} {\bf B667} (2008) 1.

\bibitem{Davies:2008sw}
{\bf HPQCD} Collaboration, C.~T.~H. Davies {\em et.~al.}, {\it {Update:
  Accurate Determinations of $\alpha_s$ from Realistic Lattice QCD}},  {\em
  Phys. Rev.} {\bf D78} (2008) 114507,
  [\href{http://xxx.lanl.gov/abs/0807.1687}{{\tt arXiv:0807.1687}}].

\bibitem{Allison:2008xk}
{\bf HPQCD} Collaboration, I.~Allison {\em et.~al.}, {\it {High-Precision
  Charm-Quark Mass from Current-Current Correlators in Lattice and Continuum
  QCD}},  {\em Phys. Rev.} {\bf D78} (2008) 054513,
  [\href{http://xxx.lanl.gov/abs/0805.2999}{{\tt arXiv:0805.2999}}].

\bibitem{Shintani:2010ph}
E.~Shintani {\em et.~al.}, {\it {Strong coupling constant from vacuum
  polarization functions in three-flavor lattice QCD with dynamical overlap
  fermions}},  \href{http://xxx.lanl.gov/abs/1002.0371}{{\tt arXiv:1002.0371}}.

\bibitem{Mason:2005zx}
{\bf HPQCD} Collaboration, Q.~Mason {\em et.~al.}, {\it {Accurate
  determinations of alpha(s) from realistic lattice QCD}},  {\em Phys. Rev.
  Lett.} {\bf 95} (2005) 052002,
  [\href{http://xxx.lanl.gov/abs/hep-lat/0503005}{{\tt hep-lat/0503005}}].

\bibitem{Shintani:pos_lat09}
E.~Shintani {\em et.~al.}, {\it Determination of $\alpha_s$ in 2+1-flavor qcd
  through valuum polarization function},  {\em
  \href{http://pos.sissa.it/archive/conferences/091/207/LAT2009_207.pdf}{PoS(L%
ATTICE 2009)207}} (2009).

\bibitem{Boucaud:2008gn}
P.~Boucaud {\em et.~al.}, {\it {Ghost-gluon running coupling, power corrections
  and the determination of $\Lambda_{\bar {\rm MS}}$}},  {\em Phys. Rev.} {\bf
  D79} (2009) 014508, [\href{http://xxx.lanl.gov/abs/0811.2059}{{\tt
  arXiv:0811.2059}}].

\bibitem{Desoto:2009aw}
F.~De~soto, M.~Gravina, O.~Pene, and J.~Rodriguez-Quintero, {\it
  {$\Lambda_{QCD}$ from gluon and ghost propagators}},
  \href{http://xxx.lanl.gov/abs/0911.4505}{{\tt arXiv:0911.4505}}.

\bibitem{Sternbeck:2010xu}
A.~Sternbeck {\em et.~al.}, {\it {QCD Lambda parameter from Landau-gauge gluon
  and ghost correlations}},  \href{http://xxx.lanl.gov/abs/1003.1585}{{\tt
  arXiv:1003.1585}}.

\bibitem{Davies:2009ih}
C.~T.~H. Davies {\em et.~al.}, {\it {Precise charm to strange mass ratio and
  light quark masses from full lattice QCD}},
  \href{http://xxx.lanl.gov/abs/0910.3102}{{\tt arXiv:0910.3102}}.

\bibitem{Mawhinney:2009jy}
{\bf RBC} Collaboration, R.~Mawhinney, {\it {NLO and NNLO chiral fits for 2+1
  flavor DWF ensembles}},  {\em PoS} {\bf LAT2009} (2009) 081,
  [\href{http://xxx.lanl.gov/abs/0910.3194}{{\tt arXiv:0910.3194}}].

\bibitem{Bazavov:2009tw}
{\bf The MILC} Collaboration, A.~Bazavov {\em et.~al.}, {\it {Results from the
  MILC collaboration's SU(3) chiral perturbation theory analysis}},  {\em PoS}
  {\bf LAT2009} (2009) 079, [\href{http://xxx.lanl.gov/abs/0910.3618}{{\tt
  arXiv:0910.3618}}].

\bibitem{Lytle:2009xm}
A.~T. Lytle, {\it {Non-perturbative calculation of $Z_m$ using Asqtad
  fermions}},  {\em PoS} {\bf LAT2009} (2009) 202,
  [\href{http://xxx.lanl.gov/abs/0910.3721}{{\tt arXiv:0910.3721}}].

\bibitem{Noaki:2009sk}
{\bf TWQCD} Collaboration, J.~Noaki {\em et.~al.}, {\it {Chiral properties of
  light mesons with $N_f=2+1$ overlap fermions}},  {\em PoS} {\bf LAT2009}
  (2009) 096, [\href{http://xxx.lanl.gov/abs/0910.5532}{{\tt
  arXiv:0910.5532}}].

\bibitem{Aoki:2009ix}
{\bf PACS-CS} Collaboration, S.~Aoki {\em et.~al.}, {\it {Physical Point
  Simulation in 2+1 Flavor Lattice QCD}},  {\em Phys. Rev.} {\bf D81} (2010)
  074503, [\href{http://xxx.lanl.gov/abs/0911.2561}{{\tt arXiv:0911.2561}}].

\bibitem{DellaMorte:2005kg}
{\bf ALPHA} Collaboration, M.~Della~Morte {\em et.~al.}, {\it {Non-perturbative
  quark mass renormalization in two-flavor QCD}},  {\em Nucl. Phys.} {\bf B729}
  (2005) 117--134, [\href{http://xxx.lanl.gov/abs/hep-lat/0507035}{{\tt
  hep-lat/0507035}}].

\bibitem{Gockeler:2006vi}
M.~Gockeler {\em et.~al.}, {\it {Simulating at realistic quark masses: Light
  quark masses}},  {\em PoS} {\bf LAT2006} (2006) 160,
  [\href{http://xxx.lanl.gov/abs/hep-lat/0610071}{{\tt hep-lat/0610071}}].

\bibitem{Scholz:2009yz}
E.~E. Scholz, {\it {Light Hadron Masses and Decay Constants}},
  \href{http://xxx.lanl.gov/abs/0911.2191}{{\tt arXiv:0911.2191}}.

\bibitem{McNeile:2009eq}
{\bf HPQCD} Collaboration, C.~McNeile {\em et.~al.}, {\it {Towards precise
  relativistic b quarks on the lattice}},
  \href{http://xxx.lanl.gov/abs/0910.2921}{{\tt arXiv:0910.2921}}.

\bibitem{Mason:2005bj}
{\bf HPQCD} Collaboration, Q.~Mason, H.~D. Trottier, R.~Horgan, C.~T.~H.
  Davies, and G.~P. Lepage, {\it {High-precision determination of the
  light-quark masses from realistic lattice QCD}},  {\em Phys. Rev.} {\bf D73}
  (2006) 114501, [\href{http://xxx.lanl.gov/abs/hep-ph/0511160}{{\tt
  hep-ph/0511160}}].

\bibitem{Dimopoulos:2009hy}
P.~Dimopoulos {\em et.~al.}, {\it {Scaling and ChPT Description of Pions from
  $N_f=2$ twisted mass QCD}},  {\em PoS} {\bf LAT2009} (2009) 117,
  [\href{http://xxx.lanl.gov/abs/0912.5198}{{\tt arXiv:0912.5198}}].

\bibitem{Baron:2009wt}
{\bf ETM} Collaboration, R.~Baron {\em et.~al.}, {\it {Light Meson Physics from
  Maximally Twisted Mass Lattice QCD}},
  \href{http://xxx.lanl.gov/abs/0911.5061}{{\tt arXiv:0911.5061}}.

\bibitem{Aoki:2006ib}
Y.~Aoki, C.~Dawson, J.~Noaki, and A.~Soni, {\it Proton decay matrix elements
  with domain-wall fermions},  {\em Phys. Rev.} {\bf D75} (2007) 014507,
  [\href{http://xxx.lanl.gov/abs/hep-lat/0607002}{{\tt hep-lat/0607002}}].

\bibitem{Gockeler:2008we}
{\bf QCDSF} Collaboration, M.~Gockeler {\em et.~al.}, {\it {Non-perturbative
  renormalization of three-quark operators}},  {\em Nucl. Phys.} {\bf B812}
  (2009) 205--242, [\href{http://xxx.lanl.gov/abs/0810.3762}{{\tt
  arXiv:0810.3762}}].

\bibitem{Braun:2008ur}
{\bf QCDSF} Collaboration, V.~M. Braun {\em et.~al.}, {\it {Nucleon
  distribution amplitudes and proton decay matrix elements on the lattice}},
  {\em Phys. Rev.} {\bf D79} (2009) 034504,
  [\href{http://xxx.lanl.gov/abs/0811.2712}{{\tt arXiv:0811.2712}}].

\bibitem{Donini:1995zd}
A.~Donini {\em et.~al.}, {\it Non-perturbative renormalization of the $\delta
  s=2$ operator and the heavy-light axial current},  {\em Nucl. Phys. Proc.
  Suppl.} {\bf 47} (1996) 489--492,
  [\href{http://xxx.lanl.gov/abs/hep-lat/9509078}{{\tt hep-lat/9509078}}].

\bibitem{ElKhadra:1996mp}
A.~X. El-Khadra, A.~S. Kronfeld, and P.~B. Mackenzie, {\it {Massive Fermions in
  Lattice Gauge Theory}},  {\em Phys. Rev.} {\bf D55} (1997) 3933--3957,
  [\href{http://xxx.lanl.gov/abs/hep-lat/9604004}{{\tt hep-lat/9604004}}].

\bibitem{Aoki:2001ra}
S.~Aoki, Y.~Kuramashi, and S.-i. Tominaga, {\it {Relativistic heavy quarks on
  the lattice}},  {\em Prog. Theor. Phys.} {\bf 109} (2003) 383--413,
  [\href{http://xxx.lanl.gov/abs/hep-lat/0107009}{{\tt hep-lat/0107009}}].

\bibitem{Christ:2006us}
N.~H. Christ, M.~Li, and H.-W. Lin, {\it {Relativistic heavy quark effective
  action}},  {\em Phys. Rev.} {\bf D76} (2007) 074505,
  [\href{http://xxx.lanl.gov/abs/hep-lat/0608006}{{\tt hep-lat/0608006}}].

\bibitem{Li:2008kb}
{\bf RBC and UKQCD} Collaboration, M.~Li, {\it {Bottom spectroscopy on
  dynamical 2+1 flavor domain wall fermion lattices with a relativistic heavy
  quark action}},  {\em PoS} {\bf LATTICE2008} (2008) 120,
  [\href{http://xxx.lanl.gov/abs/0810.0040}{{\tt arXiv:0810.0040}}].

\bibitem{Li:2007en}
M.~Li and H.-W. Lin, {\it {Charm spectroscopy on dynamical 2+1 flavor domain
  wall fermion lattices with a relativistic heavy quark action}},  {\em PoS}
  {\bf LAT2007} (2007) 117, [\href{http://xxx.lanl.gov/abs/0710.0910}{{\tt
  arXiv:0710.0910}}].

\bibitem{Skouroupathis:2010mq}
A.~Skouroupathis and H.~Panagopoulos, {\it {Two-loop renormalization of fermion
  bilinear operators on the lattice}},  {\em PoS} {\bf LAT2009} (2009) 200,
  [\href{http://xxx.lanl.gov/abs/1002.3513}{{\tt arXiv:1002.3513}}].

\bibitem{Constantinou:2010dn}
M.~Constantinou, H.~Panagopoulos, and F.~Stylianou, {\it {Perturbative
  renormalization of GPDs to $O(a^2)$, for various fermion/gluon actions}},
  \href{http://xxx.lanl.gov/abs/1001.1498}{{\tt arXiv:1001.1498}}.

\bibitem{Constantinou:2010wn}
M.~Constantinou, V.~Lubicz, H.~Panagopoulos, A.~Skouroupathis, and
  F.~Stylianou, {\it {$O(a^2)$ corrections to 1-loop matrix elements of
  4-fermion operators with improved fermion/gluon actions}},
  \href{http://xxx.lanl.gov/abs/1001.1241}{{\tt arXiv:1001.1241}}.

\bibitem{DiRenzo:2004ge}
F.~Di~Renzo and L.~Scorzato, {\it {Numerical stochastic perturbation theory for
  full QCD}},  {\em JHEP} {\bf 10} (2004) 073,
  [\href{http://xxx.lanl.gov/abs/hep-lat/0410010}{{\tt hep-lat/0410010}}].

\bibitem{Brambilla:2010qb}
M.~Brambilla, F.~Di~Renzo, and L.~Scorzato, {\it {High loop renormalization
  constants for Wilson fermions/Symanzik improved gauge action}},  {\em PoS}
  {\bf LAT2009} (2009) 211, [\href{http://xxx.lanl.gov/abs/1002.0446}{{\tt
  arXiv:1002.0446}}].

\bibitem{Torrero:2009kc}
C.~Torrero and G.~S. Bali, {\it {NSPT calculations in the Schrodinger
  Functional formalism}},  \href{http://xxx.lanl.gov/abs/0910.4138}{{\tt
  arXiv:0910.4138}}.

\bibitem{DiRenzo:2009ni}
F.~Di~Renzo, E.~M. Ilgenfritz, H.~Perlt, A.~Schiller, and C.~Torrero, {\it
  {Two-point functions of quenched lattice QCD in Numerical Stochastic
  Perturbation Theory. (I) The ghost propagator in Landau gauge}},  {\em Nucl.
  Phys.} {\bf B831} (2010) 262--284,
  [\href{http://xxx.lanl.gov/abs/0912.4152}{{\tt arXiv:0912.4152}}].

\bibitem{Baron:2009zq}
R.~Baron {\em et.~al.}, {\it {First results of ETMC simulations with Nf=2+1+1
  maximally twisted mass fermions}},
  \href{http://xxx.lanl.gov/abs/0911.5244}{{\tt arXiv:0911.5244}}.

\bibitem{Bazavov:2009wm}
{\bf MILC} Collaboration, A.~Bazavov {\em et.~al.}, {\it {Progress on four
  flavor QCD with the HISQ action}},  {\em PoS} {\bf LAT2009} (2009) 123,
  [\href{http://xxx.lanl.gov/abs/0911.0869}{{\tt arXiv:0911.0869}}].

\end{thebibliography}\endgroup

\end{document}